\documentclass[twocolumn,aps,prl,superscriptaddress,footinbib,amsmath,amssymb,floatfix]{revtex4-1}
\usepackage{physics}
\usepackage[usenames,dvipsnames]{color}
\usepackage{hyperref}
\hypersetup{colorlinks,
  citecolor=Blue,
  linkcolor=Blue,
  urlcolor=Blue
}
\usepackage{bm}
\usepackage{graphicx}
\usepackage{dsfont} 
\usepackage{geometry}
\usepackage{cancel} 
\usepackage{math tools} 
\usepackage{wasysym}
\usepackage{siunitx} %this allows \SI{}{} specifications of units

\renewcommand{\Re}{\operatorname{Re}}
\renewcommand{\Im}{\operatorname{Im}}
\newcommand{\ra}{\rangle}
\newcommand{\la}{\langle}
\newcommand{\bk}{{\bm{k}}}
\newcommand{\sab}{\sigma_{\alpha \beta} }

\begin{document}

\title{Conductivity spectrum of ultracold atoms in an optical lattice}

\author{Rhys Anderson}
\author{Fudong Wang}
\author{Peihang Xu}
\author{Vijin Venu}
\author{Stefan Trotzky}
\affiliation{Department of Physics, University of Toronto, Ontario M5S 1A7 Canada}
\author{Fr\'ed\'eric Chevy}
\affiliation{Laboratoire Kastler Brossel, ENS-PSL Research University, CNRS, UPMC-Sorbonne Universit\'e, Coll\`ege de France}
\author{Joseph H. Thywissen}
\affiliation{Department of Physics, University of Toronto, Ontario M5S 1A7 Canada}
\affiliation{Canadian Institute for Advanced Research, Toronto, Ontario M5G~1M1 Canada}

% ------------------------------- Abstract ------------------------------ 
\begin{abstract}
We measure the conductivity of neutral fermions in a cubic optical lattice. Using in-situ fluorescence microscopy, we observe the alternating current resultant from a single-frequency uniform force applied by displacement of a weak harmonic trapping potential. In the linear response regime, a neutral-particle analogue of Ohm's law gives the conductivity as the ratio of total current to force. For various lattice depths, temperatures, interaction strengths, and fillings, we measure both real and imaginary conductivity, up to a frequency sufficient to capture the transport dynamics within the lowest band. The spectral width of the real conductivity reveals the current dissipation rate in the lattice, and the integrated spectral weight is related to thermodynamic properties of the system through a sum rule. The global conductivity decreases with increased band-averaged effective mass, which at high temperatures approaches a T-linear regime. Relaxation of current is observed to require a finite lattice depth, which breaks Galilean invariance and enables damping through collisions between fermions.
\end{abstract}
\maketitle

The resistance of a metal is normally dominated by phonons, impurities, and crystalline defects, with electron-electron scattering playing a minor role.
For ultra-pure samples, it has been found that this situation can be reversed, with collisional physics instead playing a major role in electrical properties \cite{Molenkamp:1994,Kivelson:2011,Bandurin:2016cp,Guo3068}.
Optical lattices provide ultracold atoms with a crystalline environment of comparable purity, and also with an effectively infinite Debye temperature \cite{Gross:2017do}. 
These properties allow for the study of transport in conditions inaccessible to typical materials: at temperatures comparable to the Fermi energy, yet where phonons are absent and the crystal remains intact \cite{Perepelitsky:2016jg}. Moreover, the strength of particle-particle scattering, which is the sole remaining source of dissipation, can be tuned using a Feshbach resonance or by adjusting the lattice depth. 

In this Letter, we study the conductivity of ultracold fermions in an optical lattice subject to weak harmonic confinement. Non-equilibrium transport of ultracold fermions in periodic potentials has been investigated previously through step response \cite{Ott:2004jk,Strohmaier:2007hw,Heinze:2013}, in a mesoscopic two-terminal geometry \cite{krinner:2015,Krinner:2017ju,Lebrat:2018}, in the context of disorder-induced localization \cite{Kondov:2015,Schreiber:2015jt}, through quasimomentum relaxation \cite{Xu:2016uv}, by observing diffusion \cite{Bakr:2018,Zwierlein:2018}, and by studying expansion dynamics \cite{Kondov:2011jd,Schneider:2012,Ronzheimer:2013,Scherg:2018}. 
Here we realize the proposal of Wu, Taylor, and Zaremba \cite{WTZ15}, closely related to the proposal of Tokuno and Giamarchi \cite{TG11}, to measure conductivity $\sigma(\omega)$ directly  through the ratio of the current response $J(\omega)$ to an alternating force $F(\omega)$. In this proposal, the weak harmonic confinement of the system results in a  low-frequency resonance in $\sigma(\omega)$ near the harmonic trap frequency. The spectral width and weight of this resonance reveal the current dissipation rate and carrier inertia, which are the key low-energy transport properties of a metal.

\begin{figure}[tb!]
\includegraphics[width=\columnwidth]{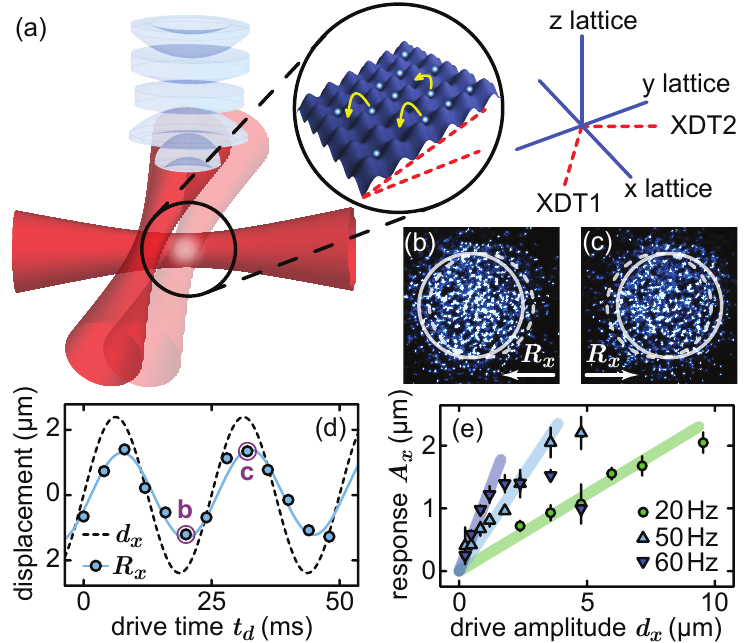}
\caption{\label{fig:Observable} {\bf Measurement}. 
(a) Atoms in a 3D optical lattice are driven by periodic displacement of one or both trapping beams (XDT). 
(b,c) In-situ images are taken after various drive times $t_d$ and the center of mass is extracted. 
(d) The displacement of the center of mass is fit to a single-frequency response (solid line) typically across two periods in trap displacement (dashed line). Data is for a 40\,Hz drive and $V=2\,E_R$ lattice. 
(e) The response amplitude $A_x$ is shown versus drive $d_x$ in typical conditions and at several $\omega$. The linear response limit is found for $A_x \lesssim \SI{1}{\micro\meter}$, as seen by comparison to the lines fit at low $A_x$.  
}
\end{figure}

Our sample is a balanced spin mixture of the two lowest hyperfine states of fermionic $^{40}$K, trapped in a cubic lattice with period $a_L=\SI{527}{\nano\meter}$. Typically $N=10^4$ atoms are loaded into the lattice at a depth $V=2.5\,E_R$ and scattering length $a_{s} = 180\,a_{0}$ between the spin states, where $E_R = \hbar^2 (\pi /a_L)^2 /2m$ is the recoil energy, $m$ is the mass of a $^{40}$K atom, and $a_{0}$ is a Bohr radius. 
At this depth, $t = h\times 570$\,Hz is the nearest-neighbor tunneling strength. Temperatures $T$ are measured independently for each dataset \cite{SM}, and typically range from $1.2\,t$ to $3.3\,t$ (here $k_\mathrm{B}=1$), or $0.3\,T_{F}$ to $0.9\,T_{F}$, where $T_F$ is the Fermi temperature.

A periodic displacement of one or both of the laser beams forming a crossed dipole trap (XDT) [Fig.~\ref{fig:Observable}(a)] creates the analogue of the voltage in an electronic conductivity measurement. 
The uniform force $F_\beta$ is linear in the trap displacement $d_\beta$ in direction $\beta$, through an in-plane spring constant $m \omega_\mathrm{XDT}^2$, where $\omega_\mathrm{XDT}=2 \pi \times 32(1)$\,Hz. The amplitude of the periodic force is increased linearly over \SI{150}{\milli\second}, and then held constant for \SI{50}{\milli\second}. After a further variable time $t_d$, up to two drive periods, the dynamics are frozen by increasing $V$ to $60\,E_R$ in \SI{0.1}{\milli\second}, and the drive is turned off. The in-situ density distribution of the cloud is recorded at $V=10^3\,E_R$ via fluorescence in a quantum gas microscope  \cite{Nelson2007,Bakr:2009bx,Sherson:2010hg,Parsons:2015ck,Haller:2015hi,Cheuk:2015jr,Omran:2015io,Mitra:2017,Ott:2016dc} 
apparatus described previously \cite{Edge:2015gr}.

From images of the central four planes [see Fig.~\ref{fig:Observable}(b,c)], we determine projections of a site-granulated centre-of-mass position operator $\hat{R}_{\alpha=x,y} =N^{-1} \sum_{i,s} r_{\alpha,i} \hat{n}_{i,s}$, where $\hat{n}_{i,s}$ is the number operator for an atom of spin $s$ on lattice site $i$ located at $r_{\alpha,i}$. $\langle \hat{R}_\alpha (t) \rangle$ is fit to $A_\alpha \cos[\omega t_d - \phi_\alpha]$ [see Fig.~\ref{fig:Observable}(d)], where $\omega$ is the drive frequency (typically $2\pi\times$\,10--200\,Hz) and $A_\alpha$ and $\phi_\alpha$ are fit parameters. The steady-state bulk current is then
$\langle \hat J _\alpha(\omega) \rangle = N d\langle \hat R_\alpha(\omega) \rangle/dt$.
As shown in Fig.~\ref{fig:Observable}(e), remaining in the linear-response regime restricts $\langle \hat R_\alpha \rangle$ to the micron scale, emphasizing the need for high-resolution measurement. In complex notation, the global conductivity $\sab(\omega)$ can be determined through the equivalent of Ohm's law \cite{WTZ15}, 
\begin{equation} \label{eq:ohm} 
\langle \hat J _\alpha(\omega) \rangle =\sum_\beta \sab(\omega) F_\beta(\omega) \,. \end{equation} 
In terms of fit variables and drive strength, $\sab(\omega) = -i N \omega A_\alpha (\omega) \exp[i \phi_\alpha (\omega)] /F_\beta (\omega)$. 
We write the conductivity in dimensionless form $\sigma/\sigma_0$, where $\sigma_0 \equiv N a_L^2 / \hbar$ sets the scale of the Mott-Ioffe-Regel limit \cite{Gunnarsson:2003fm}.

\begin{figure}[t!]
\includegraphics[width=\columnwidth]{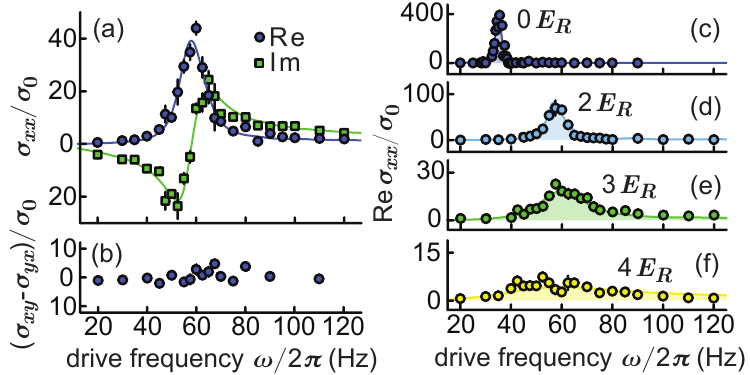}
\caption{\label{fig:sigma} {\bf Conductivity spectra}.
(a) Real and imaginary on-diagonal conductivity, and (b) difference in real off-diagonal conductivities, versus drive frequency. Here, $V = 2.5\,E_R$, with background $a_s$. Lines show fits to Eq.~(\ref{eq:conductivitymodel}). (c) For $V=0$, a Fourier-limited response is observed, with $mS_{xx}/N = 1.01(8)$ and $\Gamma=\SI{18(1)}{\per\second}$. (d,e,f) For increasing $V$, the response broadens and spectral weight (shaded) decreases. By $V=4\,E_R$, $mS_{xx}/N = 0.37(2)$ and $\Gamma=\SI{370(140)}{\per\second}$. The upwards shift in frequency is due to increased confinement from the lattice beams.}
\end{figure}

Figure~\ref{fig:sigma}(a) shows an example of on-diagonal conductivity, $\sigma_{xx}$. 
We observe both a peak in $\Re \sigma_{xx}$ and a zero-crossing in $\Im \sigma_{xx}$ at finite frequency, in contrast to the dc peak expected in a conventional metal with Drude-like response. 
This can be understood as a capacitive effect of the harmonic trap, which shifts the peak response to its oscillation frequency, renormalized by the effective mass of particles in the lattice \cite{Heinze:2013}.

Figure~\ref{fig:sigma}(b) shows that in the same conditions, ${\sigma_{xy} - \sigma_{yx}}$ is a smaller and noisy signal. When integrated using a sum rule for off-diagonal conductivity \cite{Drew:1997jb}, we find a cyclotron frequency of $2\pi \times (0 \pm 2) $\,Hz. 
This is expected, since no gauge field is applied here, however the method could be used to explore the finite-frequency anomalous conductivity of systems with broken time-reversal symmetry \cite{LeBlanc:2012co,Choi:2013eb,Trane1701207,Asteria:2018uz}. In the remainder of this Letter, we report only on-diagonal response. 

In a pure harmonic trap, $\Re \sigma_{xx}$ is a measurement-time-limited peak at the bare trap frequency \cite{Kohn:1961et} [see Fig.~\ref{fig:sigma}(c)], but the addition of a lattice broadens the response by enabling current dissipation [Figs.~\ref{fig:sigma}(d--f)]. The Kubo relation \cite{Kubo:1957cl,Mahan} gives $\sigma(\omega)$ as the Fourier transform of the retarded current-current correlation function, and thus a finite current lifetime $\tau$ broadens the ac conductivity spectra by $\tau^{-1}$. The damping of the current is seen through the diminished quality of the resonance, similar to cavity-perturbation techniques employed in microwave spectroscopy of conductors \cite{klein:1993,Hosseini:1999}.
In a nearly pure lattice structure, collisions between the fermions are responsible for dissipation of current. However, low-energy collisions in the parabolic sector of the dispersion relation $\epsilon(\bk)$ do not contribute, since velocity and quasimomenta are proportional, as in free space \cite{Kohn:1961et,Orso:2004,Rosch:2006gz,WZ14,WTZ15}. The full band structure breaks Galilean invariance at higher quasimomenta, and enables collisional current dissipation \cite{pal:2012}. 

The broadening of the global response includes not only irreversible decoherence due to collisions, but also dephasing due to dispersion. To deconvolve these two effects, we developed an effective model based on linear response theory using the exact eigenmode structure of the single-band confined-lattice Hamiltonian \cite{Hooley:2004ct,Rey:2005bw,Ott:2004dx}, 
$\hat{H}_\mathrm{CL} = \hat{H}_{0} + \hat{H}_{P}$. In this model, a tight-binding (TB) kinetic energy $\hat{H}_{0} =-t \sum_{\la i,j \ra,s} \hat{c}^\dag_{i,s} \hat{c}_{j,s}$, where $\hat{c}_{i,s}$ is the annihilation operator and $\la i,j \ra$ are adjacent sites, is combined with parabolic confinement $\hat{H}_{P} = \frac{m}{2} \sum_{\alpha,i,s} \omega_{0\alpha}^2 r_{\alpha, i}^2 \hat{n}_{i,s}$,
with trap frequency $\omega_{0\alpha}$ in the $\alpha$ direction. For non-interacting atoms, 
linear response theory predicts that the global conductivity at $\omega$ is given by
\begin{equation} \label{eq:conductivitymodel} 
\sigma_{xx}^\mathrm{(CL)} (\omega;\Gamma) = \frac{N \omega}{i \hbar} \sum_{p' \neq p} \frac{ (f_p - f_{p'}) |\! \bra{p'} \hat{R}_x \ket{p} \! |^2}{\omega-\omega_{p p'} + i \Gamma/2}, \end{equation}
where $f_p$ is the occupation of the eigenstate $\ket{p}$ of $\hat{H}_\mathrm{CL}$, and $\hbar \omega_{p p'}$ is the energetic splitting between states. The broadening $\Gamma$ represents the adiabatic ramp rate of the perturbation, but here is extended to also model weakly dissipative effects such as interaction-induced collisional damping \cite{gotze:1972,Nagaosa:2014,Allen:2015}. 
Conductivity spectra are fit to Eq.~(\ref{eq:conductivitymodel}) with variable $\Gamma$, Maxwell-Boltzmann $f_p(T)$, and a small ($\leq 2.5$\,Hz) trap frequency shift. 
Examples are shown in Fig.~\ref{fig:sigma}.

\begin{figure}[tb]
\includegraphics[width=\columnwidth]{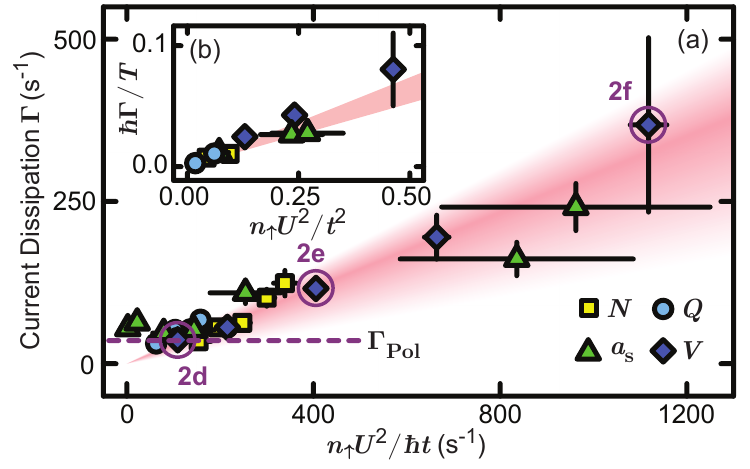}
\caption{\label{fig:tau} {\bf Transport time.} 
(a) The best-fit broadening $\Gamma$ found from $\sigma_{xx}(\omega)$ spectra in a variety of conditions. 
Varying initial $N$ from $5 \times 10^{3}$ to $50 \times 10^{3}$ creates variable density-weighted filling per spin state $n_\uparrow$, which is measured in-situ, and ranges from 0.09(1) to 0.19(2).
Varying scattering length $a_{s}$ from -240\,$a_{0}$ to 470\,$a_{0}$ results in $U/t$ that ranges from $-0.9$ to $1.8$. Depositing additional heat energy $Q$ with a non-adiabatic lattice pulse before loading results in a range of $T/t$ from 1.5 to 3.0.
The spectra in Figs.~\ref{fig:sigma}(d,e,f) correspond to the circled points in the variable-$V$ series.
The red band shows the current damping rate $\tau^{-1}$ calculated with a kinetic theory over the range of measured temperatures and densities; $\Gamma_\mathrm{Pol}$ (dashed) is measured for a non-interacting gas.
(b) The same data and theory plotted with axes non-dimensionalized to account for temperature scaling of the current dissipation rate. Some data for smallest $\Gamma$ is omitted for clarity.
}
\end{figure}

Figure~\ref{fig:tau} shows the best-fit $\Gamma$ found for $21$ different $\sigma_{xx}(\omega)$ spectra with $N$, $a_{s}$, or $V$ varied, or a quantity of heat energy $Q$ added before lattice loading to change $T$. The lowest $\Gamma$ are clustered near some non-zero minimum, which is comparable to the $\Gamma_\mathrm{Pol}=\SI{36(4)}{\per\second}$ found for a spin-polarized ensemble (dashed horizontal line) at $V=2.5\,E_R$. 
Contributors to $\Gamma_\mathrm{Pol}$ could include finite measurement time, non-quadratic terms in the confinement, and imperfections in the optical lattice. 
An independent study of the Fourier limit, with only the XDT beams \cite{SM}, gives $\Gamma \geq \SI{18(3)}{\per\second}$ for the time sequence used here.  A study of the effects of controlled disorder on $\sigma(\omega)$ is beyond the scope of this work, but would be an interesting complement to Refs.~\cite{Kondov:2015,Schreiber:2015jt,Billy:2008,Roati:2008,McKay:2008kt,Kondov:2011jd,Choi:2018}. 

For $\Gamma \gg \Gamma_\mathrm{Pol}$, Fig.~\ref{fig:tau}(a) shows a roughly linear trend versus the product $n_\uparrow U^2/t$, where $n_\uparrow$ is the density-averaged filling, and $U$ and $t$ are calculated parameters for the Hubbard $\hat H_\mathrm{FH} = \hat H_0 + \hat H_U$, with on-site interaction $\hat H_U = U \sum_i \hat{n}_{i\uparrow} \hat{n}_{i\downarrow}$.
We interpret $\Gamma$ in this limit as a measure of the current dissipation rate $\tau^{-1}$.
The linear scaling may be understood in terms of a Boltzmann-like $n_\uparrow U^{2}$ proportionality to density and scattering cross-section, which is reasonable for the low densities and moderate interaction strengths explored here \cite{Mahan,Xu:2016uv,Schneider:2012}. 
These observed trends are compared to a three-dimensional kinetic calculation of collisional current dissipation in a uniform lattice \cite{Mahan,SM}, which in the low-$n_\uparrow$, $U \ll t$ limit, gives 
\begin{equation} \label{eq:tau}
\tau^{-1} \approx n_\uparrow \frac{U^2}{\hbar t} C(T/t) \, ,
\end{equation}
where $C(T/t)$ varies between $0.11$ and $0.36$ in the range $1.2\,t\lesssim T \lesssim 3.3\,t$ considered for these data. The calculated $\tau^{-1}$ (shaded region in Fig.~\ref{fig:tau}) compares well with the measured $\Gamma$ once $\Gamma > \Gamma_\mathrm{Pol}$. 

Contained within $C(T/t)$ is the effectiveness of scattering in dissipating current, for which the role of the lattice is essential: some collisions can exchange momentum with the light field, enabling mass current to be damped. The $T$ dependence of $C(T/t)$ is quite different from the $\sqrt{T}$ scaling of collision rate in a free gas. 
A vanishing $T^2 e^{-\Delta_{UK}/T}$ Fermi-liquid signature is expected at low $T$ and $\omega$, where $\Delta_{UK}$ is the Umklapp gap \cite{Rosch:2006gz}, whereas saturation of the rate of current dissipation will occur in the high-temperature limit \cite{Perepelitsky:2016jg}. An inflection point exists between these limits \cite{SM}, leading to an approximately linear dependence of $\tau^{-1}$ on $T/t$ in the range explored here. Plotted in Fig.~\ref{fig:tau}(b) is the same data as in Fig.~\ref{fig:tau}(a) but with dimensionless axes that reflect this temperature scaling. The scaled $\hbar \Gamma/T$ data agree well with calculations of $\hbar \tau^{-1}/T$, further supporting the conclusion that we can determine transport time from our conductivity spectra. Note that since $\hbar \Gamma/T < 1$ for all measurements in Fig.~\ref{fig:tau}(b), it is reasonable to classify our system as a conventional metal, in which only a single damping time is expected \cite{Hartnoll:2015kj}. 

A second quantity deduced from the conductivity spectra is the frequency integral of $\Re \sigma$, or `f-sum'. The exact sum rule is \cite{WTZ15,SM}
\begin{equation} \label{eq:fsum} 
S^{\infty}_{\alpha \beta} \equiv \frac{2}{\pi} \int_{0}^{\infty} d\omega \Re \sigma_{\alpha\beta}(\omega) = \frac{N}{i \hbar } \left\langle \left[ \hat{R}_\alpha, \hat{J}_\beta \right] \right\rangle, \end{equation} 
where the angle brackets denote a thermal average. For any system described by $\hat{H}=\sum_{i=1}^N \hat{p}_i^2/2m+V(\bm{\hat{r}}_1,...,\bm{\hat{r}}_N)$, one can show
$S^{\infty}_{\alpha \beta} = ({N}/{m}) \delta_{\alpha,\beta}$,
independent of temperature, interaction strength, or trapping environment. We find that this sum rule is satisfied without the lattice, as shown in Fig.~\ref{fig:sigma}(c), where $S_{xx} = 1.01(8) N/m$. 

\begin{figure}[tb] \centering
\includegraphics[width=\columnwidth]{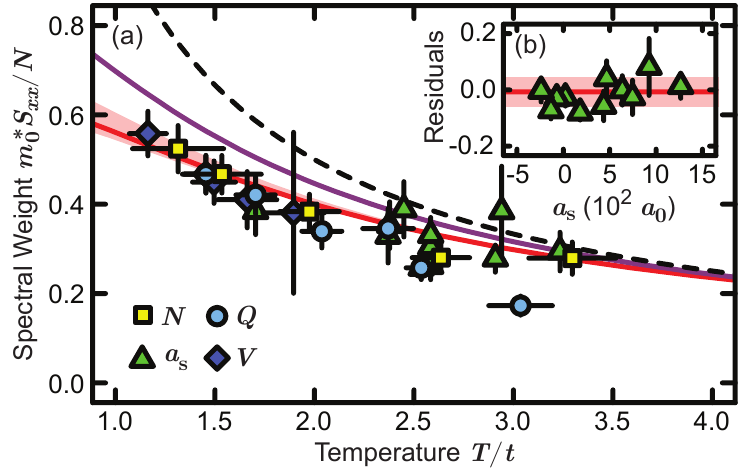}
\caption{\label{fig:fsum} {\bf Spectral weight} of data sets with various $N$, $a_s$, $Q$, and $V$, as described in the caption of Fig.~\ref{fig:tau}. 
(a) The partial f-sum $S_{xx}/N$, scaled by the TB effective mass $m^*_0$, is shown versus measured $T/t$.
Data is compared to three treatments of the uniform-lattice HM: a single band populated using MB statistics (red band), $S_{xx}^\mathrm{TB}$ (purple solid line), and the asymptotic $m^*_0 S_{xx}/N = t/T$ (black dashed line). 
The width of the red band is due to the variation of $S_{xx}$ with $V$, for fixed $T/t$. The red line is for $V = 2.5\,E_{R}$.
(b) Measured $S_{xx}$ robustly agrees with the non-interacting single-band calculation, even up to $a_s \sim 1.2\times 10^3 a_0$. 
}
\end{figure}

Figure \ref{fig:fsum} shows the spectral weight found by integration of the best-fit $\sigma_{xx}^{(\mathrm{CL})}$, across a wide range of conditions. We see that within the sampled frequency band, the spectral weight is generically less than $N/m$, 
with the remaining spectral weight transferred to inter-band transitions \cite{Heinze:2013,Reitter:2017} that need not be associated with low-frequency transport. An effective low-energy Hamiltonian $\hat{H}_{\mathrm{eff}}$ can capture this response: here, the {\em partial} f-sum is $S_{xx} = \langle [ \hat{R}_\alpha, \hat{J}^{\mathrm{sb}}_\beta ] \rangle$, with a purely single-band current $\hat{J}^{\mathrm{sb}}_\beta = [\hat{H}_\mathrm{eff}, \hat{R}_\beta]$, and its corresponding conductivity. The intra-band response has an increased carrier inertia: $S_{xx} = N/{m_\mathrm{band}}$, where $m_\mathrm{band} = \langle 1/m^*_{xx}(\bm{k}) \rangle^{-1}$ is the band mass \cite{Bari:1970bd}. Alternately, $N/{m_\mathrm{band}}$ determines the current impulse reponse to a force applied quickly compared to $\tau^{-1}$, but slowly compared to the inverse bandgap \cite{truncik:2013,Chang2014}.

The red band in Fig.~\ref{fig:fsum}(a) shows the calculated f-sum with the Fermi-Hubbard Hamiltonian, using a Maxwell-Boltzmann thermal average of single-particle states. 
Across a wide range of conditions, the data agree well with the predictions of this uniform-lattice theory with no free parameters. In the TB limit of an isotropic lattice, the f-sum is additionally a measure of kinetic energy through $S^\mathrm{TB}_{xx}=-a_L^2 \langle \hat{H}_{0} \rangle /3 \hbar^2$, and one finds
\begin{equation} \label{eq:SxxTBMB} 
S^\mathrm{TB}_{xx} = \frac{N}{m^*_0} \frac{I_1(2 t/T)}{I_0(2 t/T)} \, , 
\end{equation} 
where $I_n(z)$ is a modified Bessel function, and $m^*_0$ is the TB effective mass at $\bm{k}=0$ \cite{SM}. This result is shown as a purple line in Fig.~\ref{fig:fsum}(a), capturing the salient trend in $S_{xx}$. 

For the highest-temperature data, $m^*_0 S_{xx}/N$ approaches $t/T$ [dashed black line in Fig.~\ref{fig:fsum}(a)], which is a regime previously discussed for single-band Hubbard models in the context of $T$-linear dc resistivity \cite{Perepelitsky:2016jg,Kokalj:2017ev,Mukerjee:2006,huang:2018}.
The $1/T$ regime of spectral weight is accessible with atoms in an optical lattice because even at $T \gg t$, we do not leave the Hubbard regime, nor are phonons introduced or the crystal structure affected.
This is to be distinguished from the $T$-linear resistivity that occurs at lower temperature in both incoherent and conventional metals, and which is often attributed to the temperature dependence of $\tau^{-1}$ \cite{cao:2016,Sachdev,Zaanen:2004he,Bruin:2013hc,Hartnoll:2015kj,Zhang:2016uz,Xu:2016uv}.
Eventually, $S_{xx}$ will vanish for large $T/t$: both flat and uniformly filled bands are inert to transport.

Figure \ref{fig:fsum}(b) shows that $S_{xx}$ is relatively unaffected by $a_{s}$, despite the 
the strong variation of $\tau^{-1}$ with $a_{s}$ seen in Fig.~\ref{fig:tau}.
This illustrates a basic property of optical conductivity: scattering cannot ``destroy'' conductivity, but may only move it from one part of the $\Re \sigma (\omega)$ spectrum to another \cite{Mahan}.  

{\em Discussion ---}
The joint significance of $S_{xx}$ and $\tau$ is that their product gives an upper bound on the low-frequency conductivity. 
For example, in the weak-trap limit $\omega_0 \to 0$ of Eq.~(\ref{eq:conductivitymodel}), with fixed $\Gamma = \tau^{-1}$, the peak conductivity would be $S_{xx} \tau$. The same product is also found in Drude response $\sigma_\mathrm{D} = S_{xx}/(-i \omega + \tau^{-1})$ at $\omega=0$ \cite{mahajan:2013}. Associating conductivity with the product of a dynamical factor and a thermodynamic quantity is also found in the Nernst-Einstein form of conductivity, as the product of diffusivity and compressibility \cite{Kubo:66,Perepelitsky:2016jg,Zwierlein:2018,Bakr:2018}.

More generally, the significance of $S_{xx}$ and $\Gamma$ is that they determine the leading orders of conductivity for large $\omega$,
\begin{equation}
\sigma(\omega) \to \frac{i S_{xx}}{\omega} + \frac{\Gamma S_{xx}}{\omega^2} + \mathcal{O}\left( \frac{1}{\omega^3} \right) 
\end{equation}
up to a cutoff \cite{Chaiken,Perepelitsky:2016jg}. The first term can be shown on general grounds using Kramers-Kronig relations. The second term is model-specific, but is found in a Drude response, in our phenomenological quantum model, and in kinetic theory. Furthermore, the coefficients of these leading terms can be found from a spatial average of local responses, at least in kinetic theory \cite{SM}. This means that the f-sum and current damping rate for a trapped system can be obtained by integrating spatially the response of a uniform system. One expects such a ``local density'' picture to become valid in the high-frequency limit since the amplitude of the driven motion is vanishingly small, and over one oscillation cycle, atoms in each region of the cloud do not explore the full system. 
Experimental evidence for this correspondence is given in Figs.~\ref{fig:tau} and \ref{fig:fsum} by the excellent agreement between uniform-lattice calculations and the $S_{xx}$ and $\tau$ measured with atoms in a lattice with weak confinement. 

In sum, our work demonstrates how ac conductivity of trapped atoms can be determined, and how the spectra we observe can be understood in terms of transport time and band mass. In particular, we show that dissipation in the regime we explore is due to the combination of interactions between fermions and the breaking of Galilean invariance by the lattice. Direct extensions of this work could include measuring the conductivity spectra of strongly correlated insulators, non-Fermi-liquid metals, and resonant superfluids. 

\begin{acknowledgments}
We thank A.\ Georges, T.\ Giamarchi, A.\ MacDonald, E.\ Mueller, A.\ Paramekanti, A.\ Rosch, E.\ Taylor, Zhigang Wu, and Shizhong Zhang for stimulating conversations. This work is supported by NSERC, by AFOSR under FA9550-13-1-0063, and by ARO under W911NF-15-1-0603. F.\ C.\ acknowledges support from the European Union (ERC grants ThermoDynaMix and CritiSup2). R.\ A.\ and F.\ W.\ contributed equally.
\end{acknowledgments}

\nocite{apsrev41Control}
\bibliography{bibAnderson}

\end{document}